\documentclass[aps,prc,twocolumn,superscriptaddress]{revtex4-2}

\usepackage{amsmath,amssymb,graphicx}
\usepackage{url}
\usepackage[colorlinks,urlcolor=blue]{hyperref}
\usepackage{float}

\begin{document}

\title{The elliptic wind on jet wakes in high-energy heavy-ion collisions}

\author{Kai-Yi Wu}
\affiliation{Key Laboratory of Quark \& Lepton Physics (MOE) and Institute of Particle Physics, Central China Normal University, Wuhan 430079, China}

\author{Zhong Yang}
\email[]{zhong.yang@vanderbilt.edu}
\affiliation{Department of Physics and Astronomy, Vanderbilt University, Nashville, TN 37235}

\author{Xin-Nian Wang}
\email[]{xnwang@ccnu.edu.cn}
\affiliation{Key Laboratory of Quark \& Lepton Physics (MOE) and Institute of Particle Physics, Central China Normal University, Wuhan 430079, China}

\begin{abstract}

Energy loss by fast partons induces a Mach-cone-like medium response as they propagate inside the hot quark-gluon plasma (QGP) in high-energy heavy-ion collisions. Because the QGP is nonuniform and its initial density gradients generate collective flow, jet-induced medium response in this evolving system is also distorted by the flow and density gradient. This distortion leads to a broadened jet wake whose transverse width depends on the azimuthal angle of the jet propagation due to the elliptic anisotropy of the density gradient and the flow velocity in noncentral heavy-ion collisions. We propose and calculate the difference between the azimuth-dependent jet–hadron correlations for soft charged hadrons in in-plane and out-plane $\gamma$-jets as a measure of the elliptic broadening of the wake front and the deepening of the diffusion wake. We also study the sensitivity of this observable to the shear viscosity of the QGP. Experimental measurements of the azimuthal modulation of jet wakes induced by the wind of the elliptic flow at RHIC and LHC can provide additional constraints on the transport properties of the QGP.  

\end{abstract}

\maketitle

\noindent 1. {\bf Introduction:} Jet quenching, or suppression of high transverse momentum hadrons and jets due to parton energy loss, has been established as a powerful probe of the quark-gluon plasma (QGP) in high-energy heavy-ion collisions \cite{Bjorken:1982tu, Thoma:1990fm, Braaten:1991we, Gyulassy:1993hr, Baier:1996kr, Zakharov:1996fv, Gyulassy:1999zd, Wiedemann:2000za, Wang:2001ifa, Arnold:2002ja, Djordjevic:2006tw, Qin:2007rn,Qin:2015srf,Wang:2025lct}. The energy and momentum deposited into the hot QGP by the quenched jets will also propagate and generate a medium response that resembles a supersonic Mach-cone on a femtometer scale as shown by hydrodynamic, transport and AdS/CFT studies \cite{Casalderrey-Solana:2004fdk, Stoecker:2004qu,Ruppert:2005uz,Gubser:2007ga, Neufeld:2008fi,Qin:2009uh, Li:2010ts, Bouras:2012mh, Ayala:2016pvm, Yan:2017rku, Casalderrey-Solana:2020rsj}. 
While jet quenching probes properties of QGP at short distance scales afforded by the momentum transfer in jet-medium interaction, jet-induced medium responses are sensitive to the transport properties of QGP at the macroscopic scale of the system size \cite{Chakraborty:2006md,Cao:2020wlm, Betz:2010qh, Ma:2010dv, Tachibana:2014lja, Casalderrey-Solana:2016jvj, Tachibana:2017syd, Chen:2017zte, He:2018xjv, Zhang:2018urd, Luo:2018pto, Pablos:2019ngg, Chen:2020tbl, Tachibana:2020mtb, Chen:2021gkj,Yang:2022yfr,Du:2022oaw, Mehtar-Tani:2022zwf, Yang:2023dwc, JETSCAPE:2023hqn, Xiao:2024ffk, Li:2024pfi, Feng:2024tmc, Bossi:2024qho, Barata:2024ukm, CMS:2018jco,CMS:2018mqn, ATLAS:2020wmg, CMS:2021otx,  PHENIX:2024twd, Arslandok:2023utm, Kurkela:2026fiu}.

The medium response in general consists of the wake front behind a jet and a diffusion wake trailing the wake front. Soft partons from the original jet and radiated gluons can interact with the medium and become part of the jet-induced medium response or jet wake as often referred to.  Soft hadrons from the jet wake can lead to enhancement of jet fragmentation functions at small momentum scale \cite{Chen:2017zte,Chen:2020tbl,Chen:2021gkj} and the jet shape \cite{Casalderrey-Solana:2016jvj,Tachibana:2014lja,Tachibana:2017syd,Yang:2022nei} at large angles. The most unambiguous signal of the jet-induced medium response is the depletion of soft hadrons due to the diffusion wake in the direction opposite to the propagating jet \cite{Chen:2021gkj,Yang:2022nei}  which has been observed experimentally in $\gamma/Z$-jet \cite{ATLAS:2023fjw,ATLAS:2024prm,CMS:2024fli,CMS:2025dua} and most recently in dijet events \cite{CMS:2026mur} using the background-free rapidity asymmetry caused by diffusion wakes in jet configurations with a rapidity gap \cite{Yang:2025dqu,Yang:2025xni,Yang:2025lii}. Such precision measurements with a signal to background ratio in the order of a few $0.1\%$ usher in a new front in the study of transport properties of QGP in high-energy heavy-ion collisions.

\begin{figure}[h!]
\centering
   \includegraphics[width=0.35\textwidth]{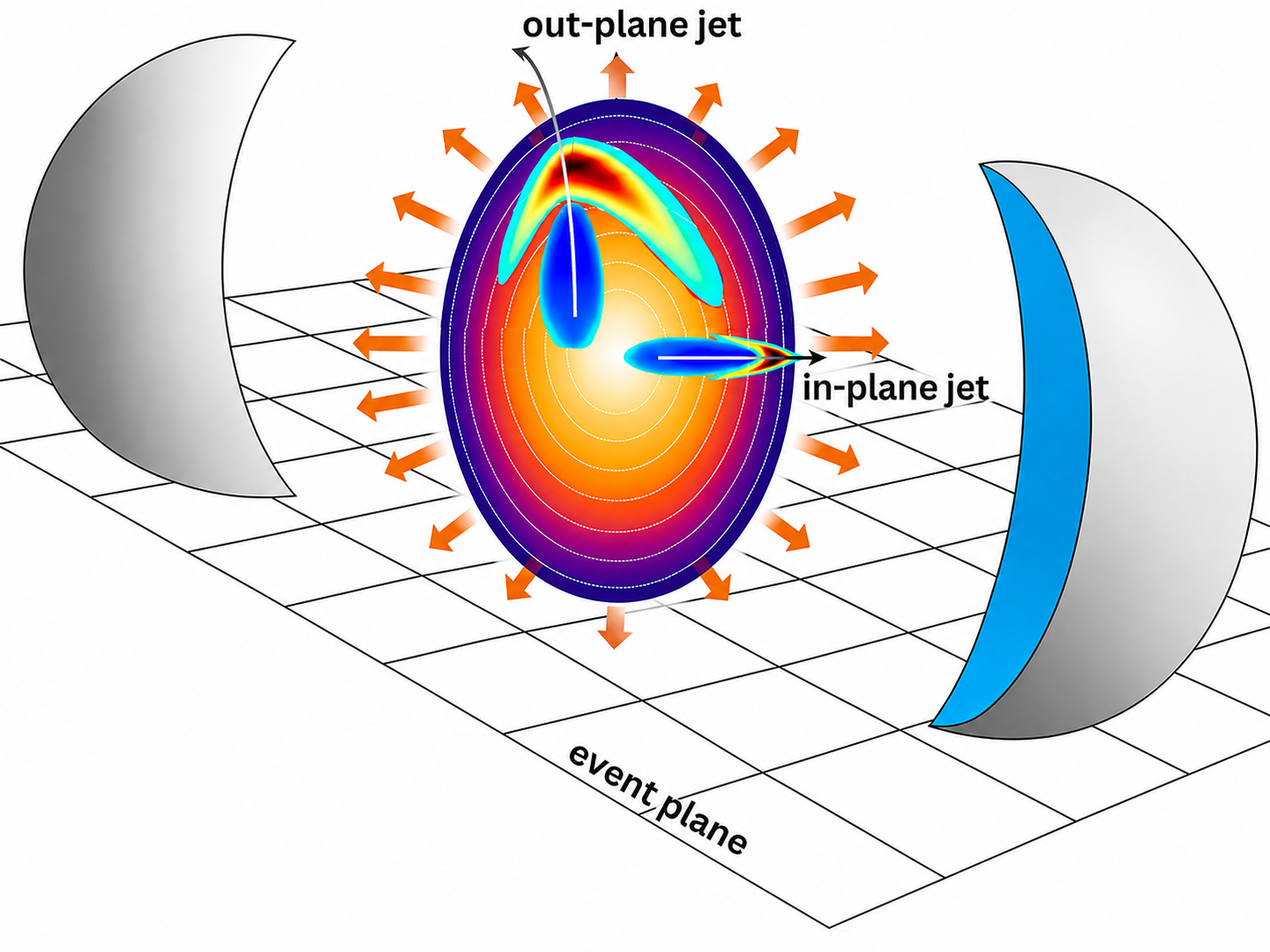}
	\caption{Illustration of the effect of elliptic flow on in-plane and out-plane jet wakes in high-energy noncentral heavy-ion collisions.}
	\label{fig:illus}
\end{figure}

The QGP fireball in high-energy heavy-ion collisions is a dynamical system that goes through rapid expansion driven by initial pressure or density gradients, leading to large collective flow in the final hadron spectra \cite{Ollitrault:1992bk,STAR:2000ekf}. It is well known that jet wakes are distorted by the density gradients and the flow velocity inside such an expanding fireball \cite{Li:2010ts,Tachibana:2020mtb,Casalderrey-Solana:2020rsj,Yang:2022yfr}, depending on the initial jet production location and propagation direction. When events are averaged over all possible initial jet production positions and propagating directions, the distortion is smeared out and loses its distinctive features in the final observed hadron spectra and correlations. 
Gradient tomography \cite{He:2020iow} and machine learning technique \cite{Yang:2022yfr,Du:2021pqa} have been proposed to select jet events with restricted initial jet production location and direction in which one can  measure the distorted medium response through trigger-hadron correlation and asymmetric jet shape \cite{Xiao:2024ffk}. In this Letter, we propose a simple measurement of the azimuthal dependence of jet-hadron correlation to study the distortion of the jet wake by the elliptic flow in noncentral heavy-ion collisions.  We first show that the transverse density gradient and flow broaden the jet wake. Because of the elliptic anisotropy of the density gradient and the radial flow, the broadening of the out-plane jet wake is larger than that of the in-plane jet wake as illustrated in Fig.~\ref{fig:illus}. We propose and calculate the difference between in-plane and out-plane jet-hadron azimuthal correlations and rapidity asymmetries in $\gamma$-jet events as measures of the distortion of the wake front and diffusion wake, respectively, by the elliptic flow. We also show the sensitivities of these new observables to the value of the specific shear viscosity within model simulations.

\noindent 2. {\bf Gradient and flow corrections to broadening:} 
In a uniform medium, the effect of flow on the medium response can be included by a boost from the local comoving frame. This is how it is often done in transport models. In a dynamic system where flow is generated by the density gradient, effects of flow and gradient are coupled and often similar. We can illustrate their influence within a Fokker-Planck (FP) diffusion equation,
\begin{equation}
\frac{\partial f}{\partial t}+\frac{\vec k_\perp}{\omega}\cdot \frac{\partial f}{\partial \vec r_\perp}=\vec\nabla_{k_\perp}\cdot(\eta_D\vec k_\perp f)+\frac{\hat q}{4}\vec\nabla_{k_\perp}^2 f(\vec k_\perp,\vec r_\perp) ,
\label{eq:diff}
\end{equation}
for parton transport  with a small transverse momentum $k_\perp/\omega\ll 1$ approximation, where $\eta_D$ is the drag constant and $\hat q$ is the jet transport coefficient defined phenomenologically as the mean transverse momentum broadening squared per unit distance. For $\eta_D=0$ and a constant jet transport coefficient $\hat q=\hat q_0$, the solution to the above diffusion equation with an initial condition $f_0(\vec k_\perp,\vec r_\perp,t=0)=(2\pi)^2\delta^2(\vec{r}_\perp)\delta^2(\vec{k}_\perp)$ is \cite{He:2020iow,Fu:2022idl,Barata:2022utc} ,
\begin{eqnarray}
f_0=3\left(\frac{4\omega}{\hat q_0 t^2}\right)^2 
\exp\left[ -(\vec{r}_\perp-\frac{\vec k_\perp}{2\omega}t)^2\frac{12\omega^2}{\hat q_0 t^3}-\frac{k_\perp^2}{\hat q_0 t}
\right],
\label{eq:solution}
\end{eqnarray}
with the transverse momentum distribution,
\begin{eqnarray}
    f_0(\vec k_\perp,t)=\int d^2r_\perp f_0(\vec k_\perp, \vec r_\perp, t)
    =\frac{4\pi}{\hat q_0 t} 
\exp\left(-\frac{k_\perp^2}{\hat q_0 t}\right).
\end{eqnarray}
For a medium with a small spatial linear gradient $\hat q=\hat q_0(1+\vec d\cdot \vec r_\perp)$ and a small flow velocity $\hat q\approx\hat q_0(1-\vec v_\perp\cdot\vec k_\perp/\omega)$, the gradient and flow corrections to the distribution up to the linear terms are $f=f_0+\delta f_d+\delta f_v$,
\begin{eqnarray}
 \label{eq:dfd}
  \delta f_d &=&t\frac{\vec d\cdot  \vec k_\perp}{3\omega}  \left(\frac{k_\perp^2}{2\hat q_0t}-1 \right)f_0(\vec k_\perp,t) +{\cal O}(d^2), \\
   \label{eq:dfv}
  \delta f_v &=&- \frac{\vec v_\perp\cdot k_\perp}{\omega}\frac{k_\perp^2}{2\hat q_0 t} f_0(\vec k_\perp,t)+{\cal O}(v_\perp^2).
\end{eqnarray}
Both the first and second moments of the gradient correction vanish. The net transverse momentum shift can be estimated by the third moments $\langle\vec K_{d,v}\rangle\equiv \langle \vec k_\perp k_\perp^2\rangle_{d,v}/(\hat q_0 t)$:
\begin{equation}
    \langle\vec K_{\perp d}\rangle=\frac{1}{6}\frac{\hat q_0 t} {\omega}\,t\vec d;\,\, \langle\vec K_{\perp v}\rangle=-\frac{3}{2}\,\frac{\hat q_0 t}{\omega}\,\vec v_\perp. 
    \label{eq:shift}
\end{equation}
If one includes the drag term in the FP diffusion equation, the results remain the same in the short time limit. Since the flow velocity is related to the linear gradient,
 $\vec v_\perp = -t \vec d/4 +{\cal O}(t^3,d^2)$
in a non-interacting QGP according to the ideal hydrodynamic equation, the gradient and flow therefore generate similar transverse-momentum shifts of a propagating parton in the direction of the gradient or toward the dense region of the medium, thereby leading to the transverse broadening of the jet-induced medium response. This is consistent with the transport simulations \cite{Chen:2021gkj}  which show that soft hadrons in jets with azimuthally asymmetrical jet shapes are anti-correlated with hard particles which are deflected away from the dense region of the QGP \cite{Xiao:2024ffk}. In noncentral heavy-ion collisions, the transverse density gradient experienced by a parton propagating in the out-plane direction is larger than that in the in-plane direction. One should therefore observe larger transverse broadening of the jet wake induced by an out-plane propagating jet than an in-plane one, as illustrated in Fig.~\ref{fig:illus}. The same gradient difference also leads to the elliptic flow of the medium, generating the elliptic wind to the jet wake in noncentral heavy-ion collisions.

\noindent 3. {\bf Jet wakes in CoLBT-hydro model simulations:} 
To verify the elliptic distortion of the jet wake and make quantitative predictions of its effects in the final hadron spectra and correlations, we employ the coupled linear Boltzmann transport and hydrodynamic (CoLBT-hydro) model \cite{Chen:2020tbl, Chen:2017zte, Zhao:2021vmu} to simulate $\gamma$-jet events in 30--50\% Pb+Pb collisions at $\sqrt{s_{\mathrm{NN}}}=5.02$~TeV. In this framework, the propagation of energetic jet shower partons and thermal recoil partons is described by the Linear Boltzmann Transport (LBT) model \cite{He:2015pra,Cao:2016gvr,Luo:2023nsi}, which includes both elastic and inelastic interactions between energetic partons and thermal medium partons. During each time step of the LBT transport, soft partons below a cutoff in the local comoving frame in the final state of each parton-medium interaction and ``negative" partons from the back-reaction are considered thermalized and contribute to a source term in the (3+1)D CCNU-LBNL viscous hydrodynamic (CLVisc) model \cite{Pang:2012he, Pang:2014ipa, Pang:2018zzo, Wu:2021fjf}  for the evolution of the bulk medium. This coupled approach thereby provides a concurrent and self-consistent description of jet transport, jet wake and the underlying bulk QGP medium including minijets from multiple parton interaction (MPI) and soft partons from the initial-state radiation (ISR). 

The initial conditions for the hydrodynamic evolution are generated with the 2D T$_\mathrm{R}$ENTo model \cite{Moreland:2014oya} with a plateau envelope in spatial rapidity. The same profile is used to sample the transverse distribution of the initial $\gamma$-jet production points. The partonic configurations of the initial $\gamma$-jet showers are generated with $|\eta_{\gamma}|<1.44$ and $p_T^{\gamma}>100$~GeV/$c$. Hadron spectra from jets and the medium response are obtained by subtracting the background spectra from hydrodynamic calculation with the same initial conditions but without the initial jets. Final jets are reconstructed from final-state hadrons 
using the FASTJET~\cite{Cacciari:2011ma} with the anti-$k_T$ algorithm and a cone size $R=0.4$, for $|\eta_{\mathrm{jet}}|<1.6$, and $p_T^{\mathrm{jet}}>50$~GeV/$c$. In this study, 30000 $\gamma$-jet events are simulated in 30-50\% Pb+Pb collisions at $\sqrt{s_{\mathrm{NN}}}=5.02$ TeV. For each value of the specific shear viscosity $\eta/s$, 2000 T$_\mathrm{R}$ENTo initial conditions are generated for CLVisc hydrodynamics and recycled across the $\gamma$-jet events.

\begin{figure}[H]
\centering
   \includegraphics[width=0.5\textwidth]{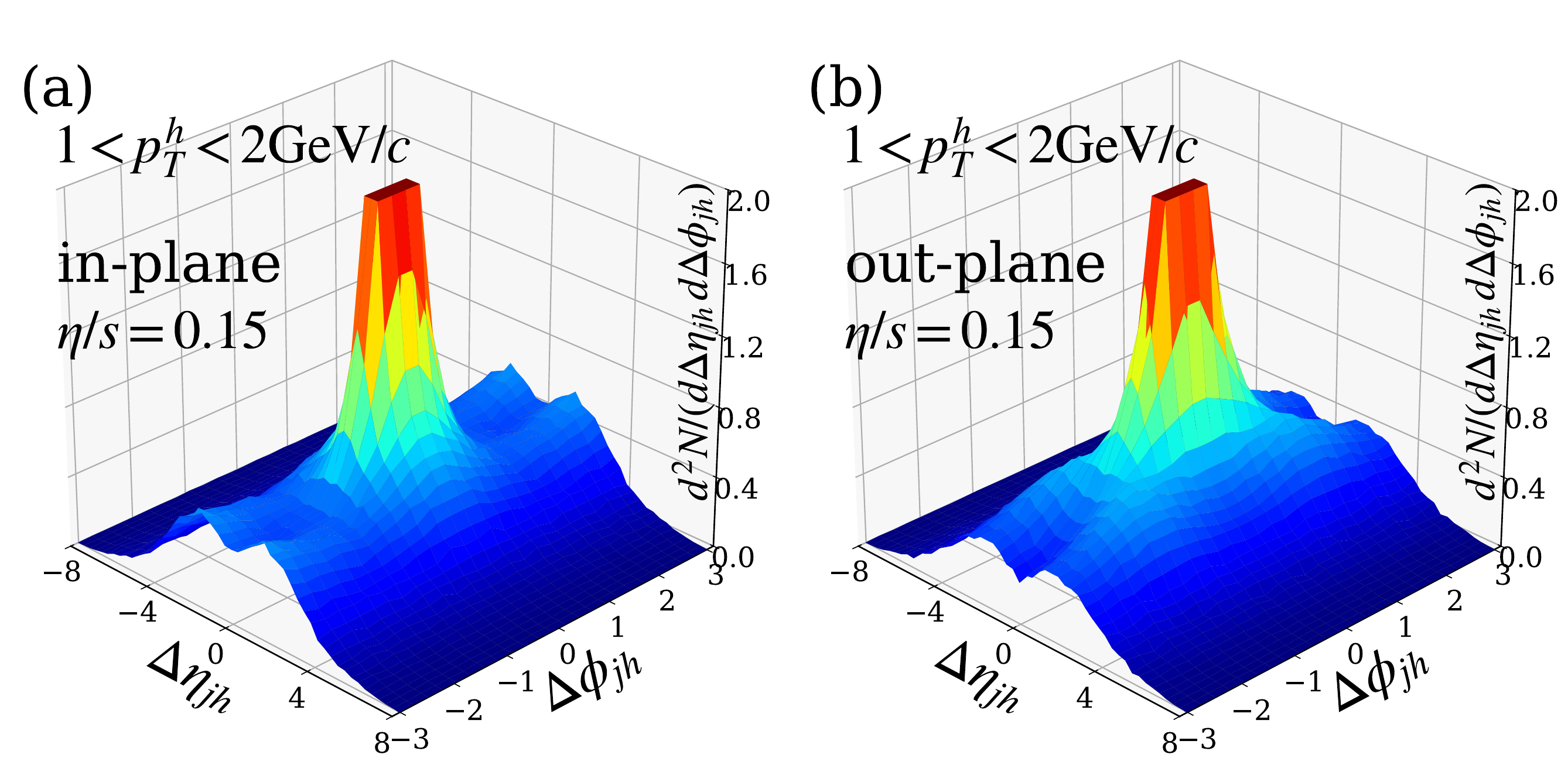}
	\caption{Jet-charged-hadron correlations for (a) in-plane and (b) out-plane $\gamma$-jets in 30-50\% Pb+Pb collisions at $\sqrt{s}=5.02$ TeV from CoLBT-hydro simulations. Values of the correlation larger than 2 are truncated for visualization.}
	\label{fig:3dcorr}
\end{figure}

Given the second-order event-plane angle $\Psi_{\rm 2}$ in each event, we classify in-plane $\gamma$-jets as those with $|\phi_{\rm jet}-\Psi_{\rm 2}|<\pi/8$ or $|\phi_{\rm jet}-\Psi_{\rm 2}+\pi|<\pi/8$ and out-plane $\gamma$-jets as  those with $| \phi_{\rm jet}-\Psi_{\rm 2}+\pi/2|<\pi/8$ or $| \phi_{\rm jet}-\Psi_{\rm 2}+3\pi/2|<\pi/8$. Fig.~\ref{fig:3dcorr} shows jet-hadron correlations as functions of rapidity $\Delta\eta_{\rm jh}\equiv\eta_{\rm h}-\eta_{\rm jet}$ and azimuthal angle $\Delta\phi_{\rm jh}\equiv\phi_{\rm h}-\phi_{\rm jet}$ for charged hadrons with $1<p_T^h<2$ GeV/$c$ for in-plane (a) and out-plane (b) $\gamma$-jets. We see a peak in the jet direction which is broadened and enhanced relative to p+p collisions \cite{Yang:2022nei} on top of the MPI-ISR ridge along $\Delta\phi_{\rm jh}$. We also observe a rapidity valley, superimposed on the MPI–ISR ridge, in the $\gamma$ direction due to the diffusion wake. These are the two typical features of the jet-induced medium response in high-energy heavy-ion collisions. One can clearly see that both the broadening of the jet peak and the diffusion-wake valley are different for in-plane and out-plane $\gamma$-jets. The underlying MPI-ISR contributions also have elliptic anisotropy, as is clearly seen at large $\Delta\eta_{\rm jh}$ where contributions from jets and medium response are small.

\begin{figure}[H]
\centering
   \includegraphics[width=0.4\textwidth]{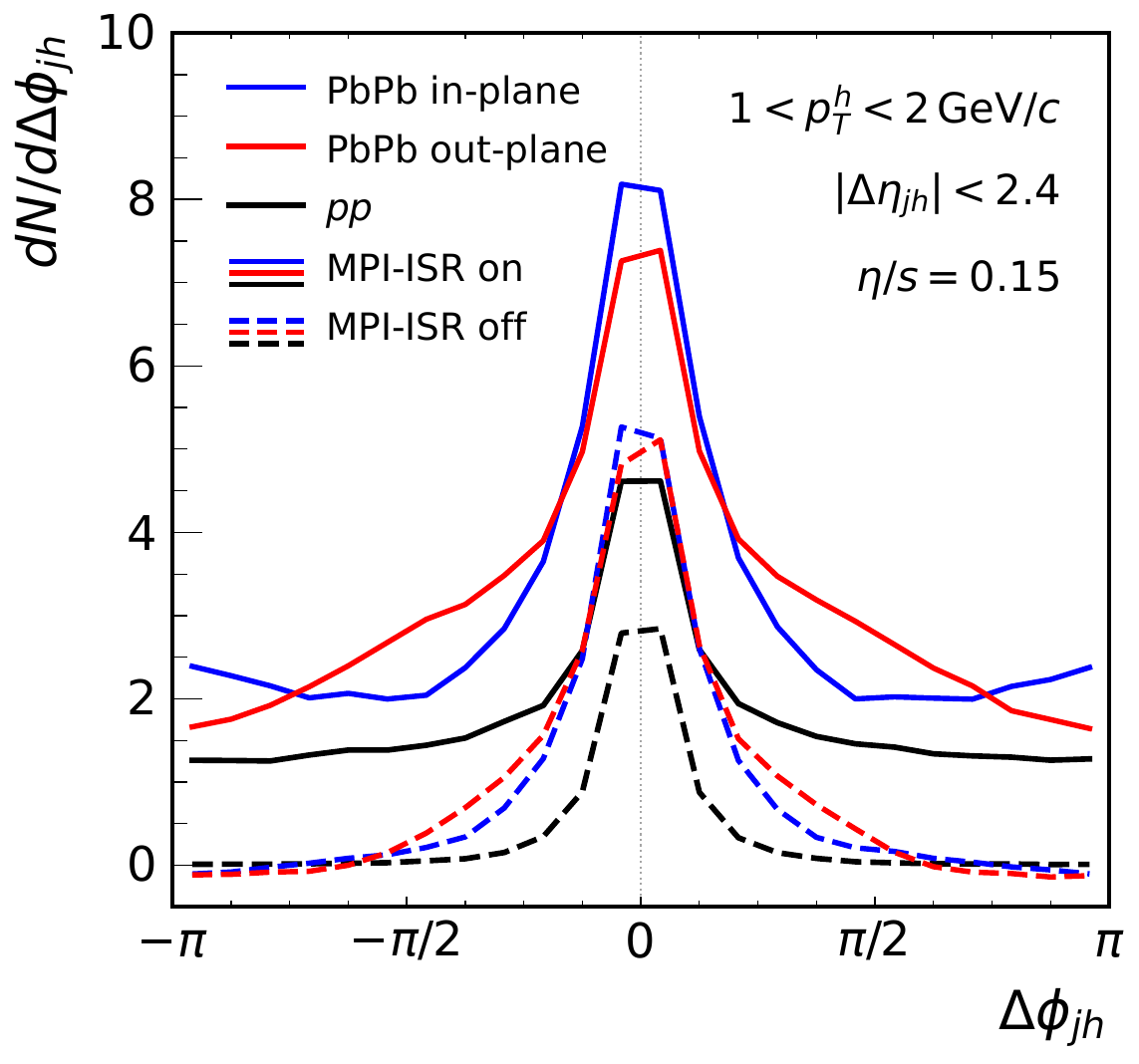}
	\caption{Jet-charged-hadron correlations in azimuthal angle for in-plane (blue) and out-plane (red) $\gamma$-jets (solid lines) in 30-50\% Pb+Pb collisions at $\sqrt{s}=5.02$ TeV as compared to that in p+p (black solid line). Results with the MPI-ISR switched off are shown in dashed lines. }
	\label{fig:dndphi1}
\end{figure}

\noindent 4. {\bf Elliptic broadening of jet wake:} 
To investigate the azimuthal dependence of jet wake  broadening due to elliptic flow and the anisotropic density gradient in detail, we project the 2D jet-hadron correlations onto the azimuthal angle $\Delta\phi_{\rm jh}$ within $|\Delta\eta_{\rm jh}|<2.4$ for in-plane (blue) and out-plane (red) $\gamma$-jets in 30-50\% Pb+Pb  as well as in p+p (black) collisions in Fig.~\ref{fig:dndphi1}. Results with the MPI-ISR switched off are shown as dashed lines.  

Without MPI and ISR, the jet-hadron correlations come only from jets and the medium response. They are clearly enhanced and broadened in Pb+Pb relative to those in p+p collisions within $|\Delta\phi_{\rm jh}|<3\pi/4$. The broadening for out-plane $\gamma$-jets is larger than in-plane ones due to elliptic flow. Because of this elliptic broadening, the enhancement of the jet-hadron correlation within $|\Delta\phi_{\rm jh}|\le \pi/8$ for out-plane $\gamma$-jets becomes slightly smaller than that for in-plane jets. This ordering of the soft hadron enhancement is opposite to what one would naively expect of the jet energy loss picture: the larger jet energy loss of out-plane jets should lead to a larger enhancement of soft hadrons than in-plane jets within the jet-cone. The elliptic broadening of the jet wake reverses this naive ordering. The integrated soft-hadron yield over a wide region $|\Delta\phi_{\rm jh}|<3\pi/4$ for out-plane jets, however, is larger than for in-plane jets, consistent with the energy loss picture.

Soft hadrons from MPI-ISR contribute to the jet-hadron correlations (solid lines) at all azimuthal angles, especially in the $\gamma$ direction. Compared to p+p collisions, MPI-ISR contributions are enhanced in Pb+Pb because of the quenching and thermalization of the associated minijets and partons from ISR. Since thermalized MPI-ISR partons  flow with the rest of the medium, their contributions to the jet-hadron correlations should have azimuthal modulations that are similar to the anisotropic flow of the bulk medium. 
The anisotropic flow of MPI-ISR hadrons leads to the different $\Delta\phi_{\rm jh}$ dependence of jet-hadron correlations in the region of the diffusion wake where the jet-hadron correlations for in-plane and out-plane jets cross each other at $|\phi_{\rm jh}|\approx 3\pi/4$.  These distinct behaviors in this region allow one to separate the jet wake and MPI-ISR contributions to the jet-hadron correlations.


\begin{figure}[H]
    \centering
    \includegraphics[width=0.48\textwidth]{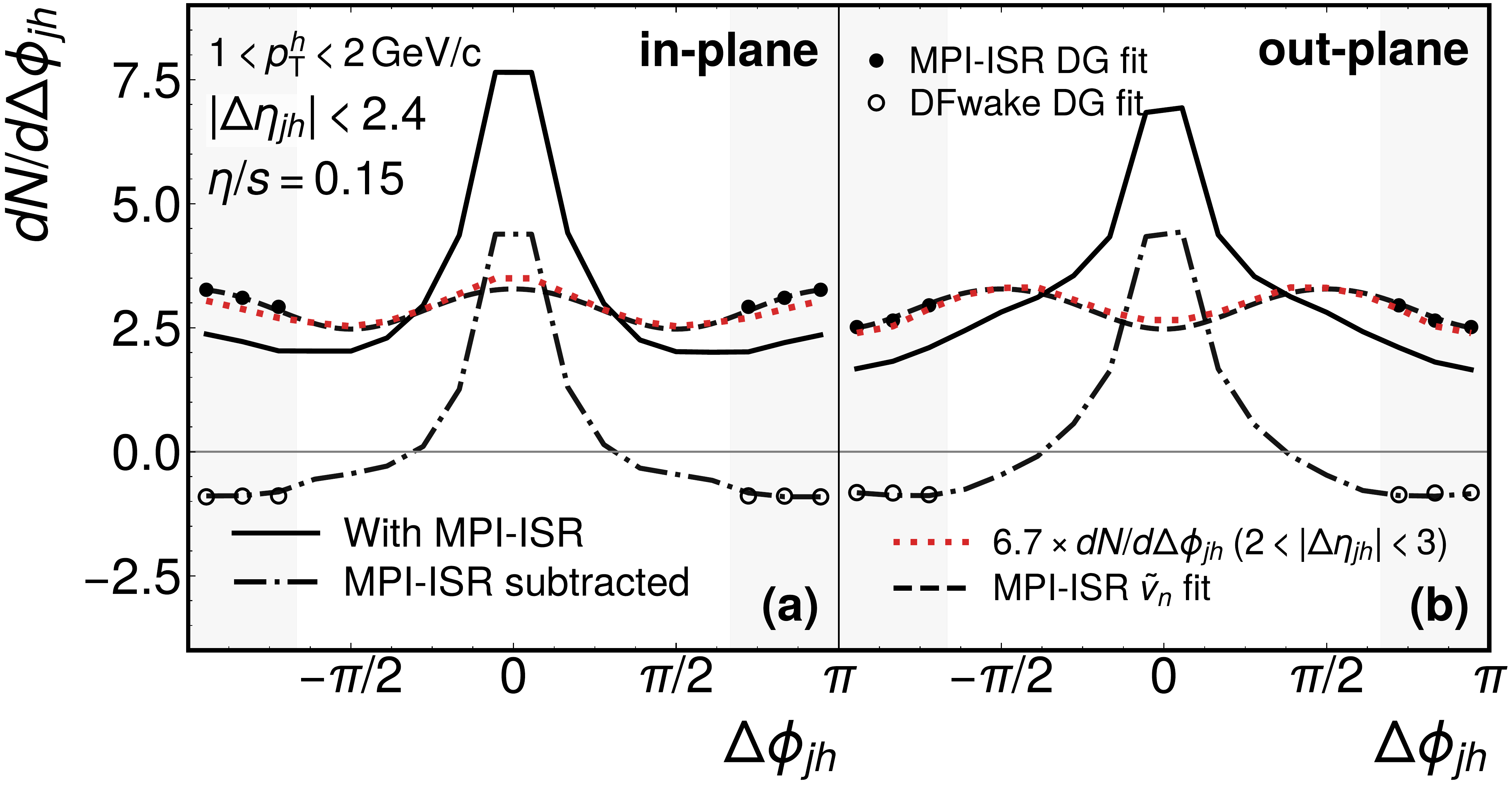}
    \caption{Diffusion wake (open circles) and MPI-ISR (filled circles) contributions to jet-hadron correlations, 
    extracted using double-Gaussian fits of the jet-hadron correlation in rapidity $\Delta\eta_{\rm jh}$. Jet-hadron correlations in $\phi_{\rm jh}$ before (solid) and after (dot-dashed) subtraction of MPI-ISR contributions, which are fitted to the anisotropic flow of the underlying events (dashed) or the jet-hadron correlation at large rapidity gap $\Delta\eta_{\rm jh}=2.5\pm 0.5$ (dotted), for (a) in-plane and (b) out-plane $\gamma$-jets in 30-50\% Pb+Pb collisions at $\sqrt{s}=5.02$ TeV.}
    \label{fig:two-gauss}
\end{figure}

To phenomenologically separate the MPI-ISR ridge and the diffusion wake valley in the jet-hadron correlations, we follow the procedure in Ref.~\cite{Yang:2022nei} and use a double-Gaussian (DG) fit of the jet-hadron correlation in rapidity in each of the six $\Delta\phi_{\rm jh}$ bins in the region $|\Delta\phi_{\rm jh}|>2\pi/3$  in the $\gamma$ direction where signals of the diffusion wake are the strongest as shown by open and closed circles in Fig.~\ref{fig:two-gauss}. We assume the azimuthal modulation of the MPI-ISR contribution is independent of the rapidity and is the same as the anisotropic flow of the underlying events without a jet, which can be fitted to
\begin{eqnarray}
\frac{dN_{\rm MPI-ISR}}{d\Delta\phi_{\rm jh}} &\approx& N_{\rm MPI-ISR}[1+2\sum_{n=2,4}\tilde v_n
\cos{n(\Delta\phi_{\rm jh}-\Psi_2)}], \nonumber \\
\tilde v_n&=&\frac{1}{n\Delta_{\rm jet}}\sin{n\Delta_{\rm jet}}\langle v_n \cos{n(\Psi_n-\Psi_2)}\rangle,
\end{eqnarray}
for in-plane jets within $|\phi_{\rm jet}-\Psi_2|<\Delta_{\rm jet}=\pi/8$ ($\Psi_2\rightarrow \Psi_2+\pi/2$ for out-plane jets) as shown by the dashed lines in Fig.~{\ref{fig:two-gauss}}. We only consider correlation between $\Psi_4$ and $\Psi_2$ event planes. Alternatively, one can also use the jet-hadron correlation at large rapidity $|\Delta\eta_{\rm jh}|=2.5\pm 0.5$, where MPI-ISR contributions dominate, to fit the MPI-ISR contributions from the DG fits as shown by the dotted lines. These two fits are very similar, indicating that the anisotropic flow of the MPI-ISR contributions is the same as the bulk medium. After the subtraction of MPI-ISR contributions, the jet-hadron correlation from the medium response is indeed broader, and the diffusion wake is deeper for out-plane jets than in-plane jets.

\begin{figure}[H]
    \centering
    \includegraphics[width=0.40\textwidth]{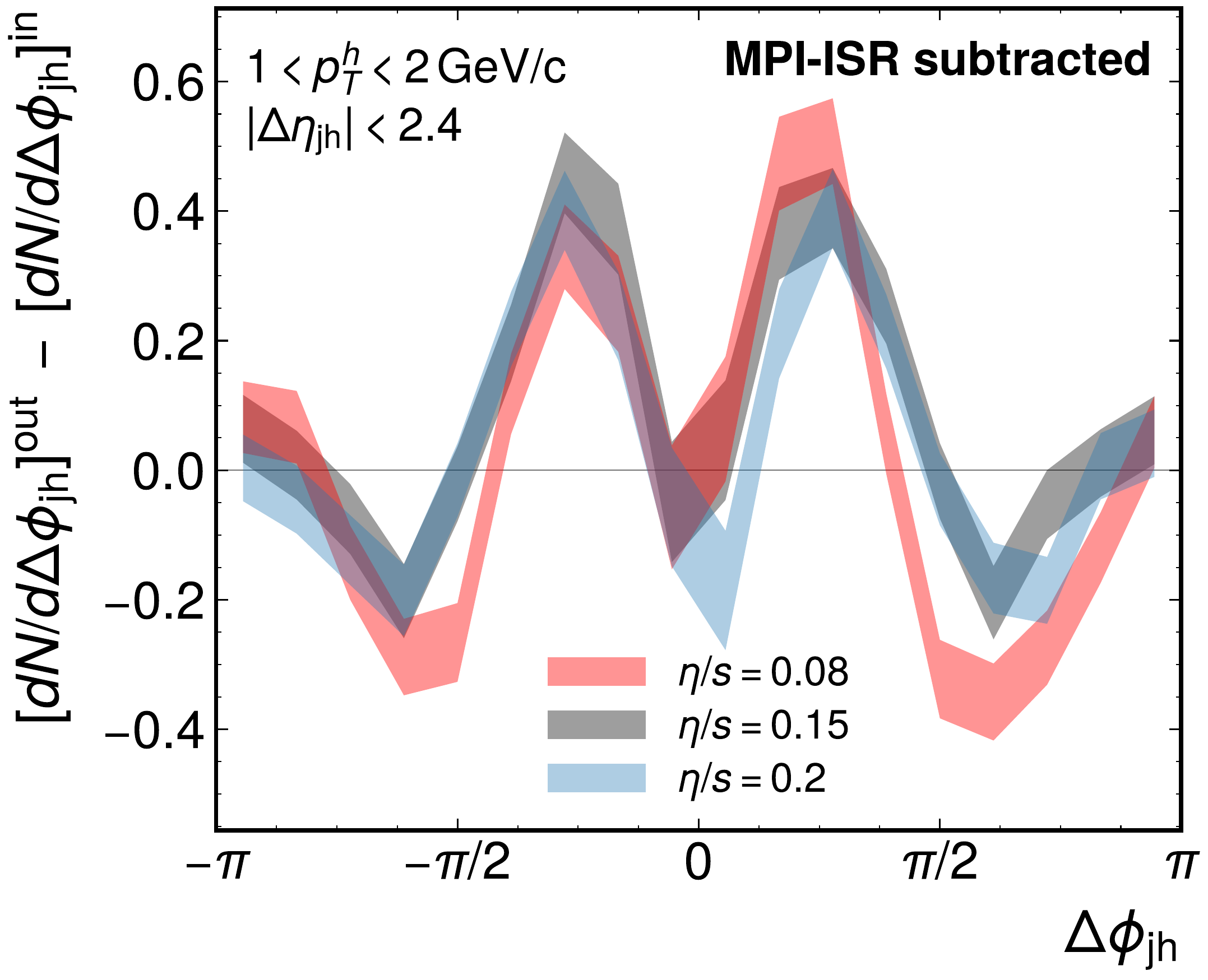}
    \caption{Difference between the out-plane and in-plane jet-hadron correlations from $\gamma$-jet induced medium response in 30-50\% Pb+Pb collisions at $\sqrt{s}=5.02$ TeV with three different values of $\eta/s$ in CLVisc hydrodynamics. }
    \label{fig:in-out-diff}
\end{figure}

To characterize the elliptic nature of the jet wake broadening, we plot the difference between the in-plane and out-plane jet wake contributions to the jet-hadron correlations in Fig~\ref{fig:in-out-diff}. Because of the elliptic broadening of the jet wake, the out-plane jet-hadron correlation in $|\Delta\phi_{\rm jh}|<\pi/2$ is enhanced relative to the in-plane jet-hadron correlation. We also show results for different values of $\eta/s=0.08$, 0.15 and 0.20. The elliptic broadening of the jet wake is reduced for smaller value of $\eta/s$ when diffusion wake becomes deeper.

\begin{figure}[H]
    \centering
    \includegraphics[width=0.40\textwidth]{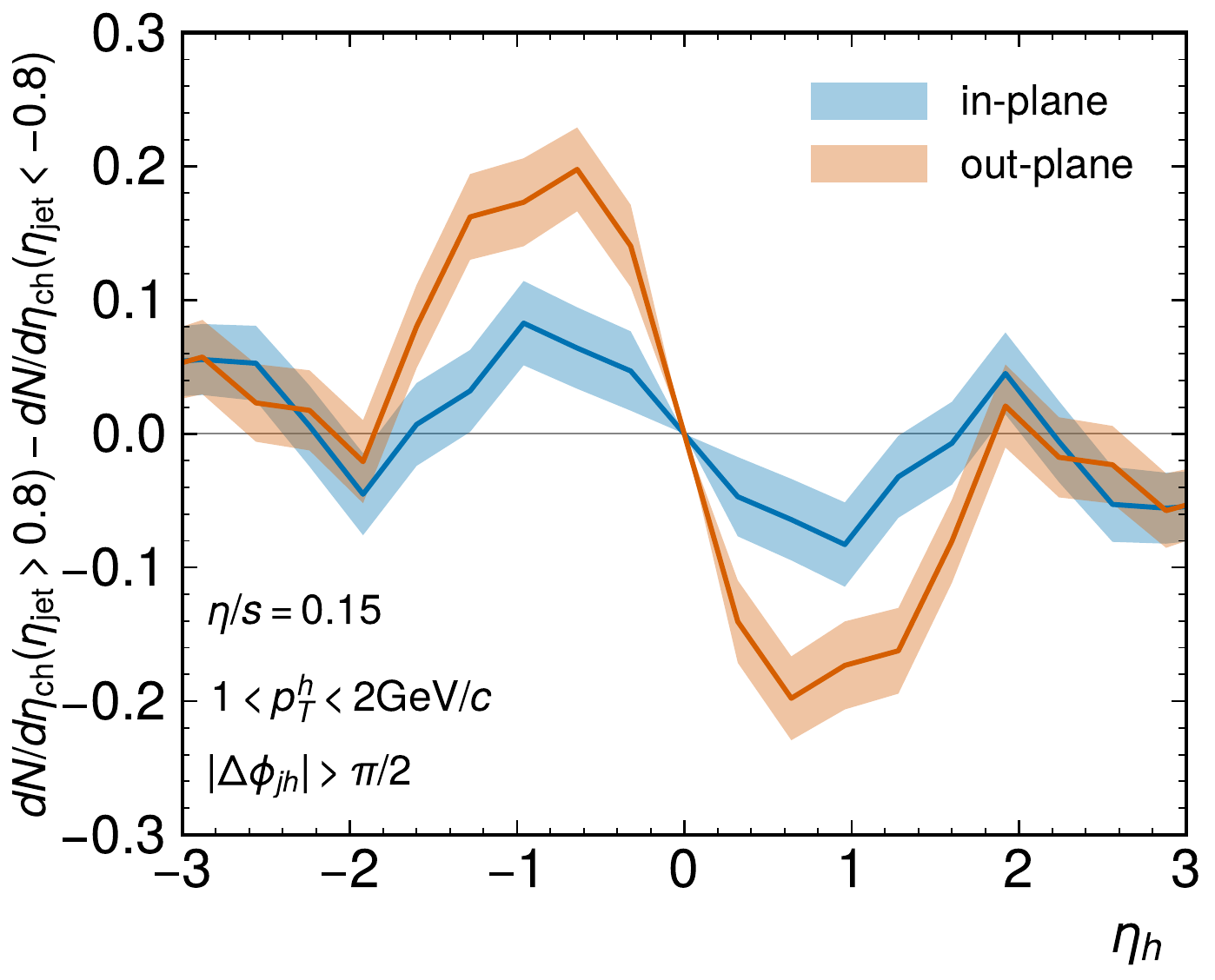}
    \caption{(Antisymmetrized rapidity asymmetry of jet-hadron correlation from the diffusion wake induced by in-plane and out-plane $\gamma$-jets in 30-50\% Pb+Pb collisions at $\sqrt{s}=5.02$ TeV.}
    \label{fig:rapasy}
\end{figure}

\noindent 5. {\bf Rapidity asymmetry from diffusion wake:} To further investigate the azimuthal dependence of the diffusion wake, we show in Fig.~\ref{fig:rapasy} the rapidity asymmetries of the in-plane and out-plane jet-hadron correlations in the $\gamma$ direction, defined as
\begin{equation}
    A(\eta_{\rm h})=\frac{dN_{\rm ch}}{d\eta_{\rm h}}|_{\eta_{\rm jet}>0.8}-\frac{dN_{\rm ch}}{d\eta_{\rm h}}|_{\eta_{\rm jet}<-0.8},
\end{equation}
for $|\Delta\phi_{\rm jh}|>\pi/2$. As proposed recently \cite{Yang:2025dqu,Yang:2025xni,Yang:2025lii}, this rapidity asymmetry of the jet-hadron correlation measures the depletion of soft hadrons caused by the diffusion wake of the propagating jet and is free of the QGP background and MPI-ISR contributions. Since the path-length-dependent jet energy loss for out-plane jets is larger than that for in-plane jets, the out-plane diffusion wake is much deeper than the in-plane one, similar to the behavior  seen in Figs.~\ref{fig:two-gauss} and \ref{fig:in-out-diff}.

\noindent 6. {\bf Conclusions:} 
We have shown that the density gradients and the resulting radial flow in a hot QGP medium can lead to the distortion of jet-induced Mach-cone-like medium response in high-energy heavy-ion collisions. The distortion can lead to the broadening of the jet wake, which should depend on the azimuthal angle of the jet propagation relative to the event plane of a noncentral collision. We propose and calculate the differences between the azimuthal distributions and rapidity asymmetries of the jet-hadron correlations for out-plane and in-plane $\gamma$-jets as measures of the broadening of jet wake by the elliptic wind in noncentral Pb+Pb collisions. We also show the sensitivities of these observables to the shear viscosity of the QGP medium. Dependence on the equation of state (EoS) of the interacting matter can also be studied.  With the proposed procedure to subtract the MPI-ISR contributions to the jet-hadron correlation which is also shown to flow with the QGP medium, experimental measurements of the elliptic broadening are feasible and can provide additional constraints on the transport properties of the QGP. 

\noindent {\bf Acknowledgment:}  This work is supported in part by NSFC under Grant No. 12535010. We thank W. Ke for helpful discussions. Computations in this study were performed at the NSC3/CCNU and NERSC under the award NP-ERCAP0032607.


\bibliographystyle{apsrev4-2}
\bibliography{Refsdw}

\appendix
\section*{Supplemental materials}

\noindent {\bf 1: Gradient and flow corrections to solutions of the FP diffusion equation:}
Consider a jet transport coefficient with a small linear spatial gradient,
$$
\hat q=\hat q_0(1+\vec d\cdot\vec x);
$$
and a small flow velocity 
$$
\hat q=\hat q_0\frac{k\cdot u}{\omega}\approx (1-\vec v_\perp\cdot\vec k_\perp/\omega).
$$
Corrections to the solution of the FP diffusion equation in Eq.~(\ref{eq:solution}) up to the linear term in $\vec d$ and $\vec v_\perp$ are,

\begin{eqnarray}
\delta f_d&=&f_0(\vec k_\perp,\vec r_\perp,t)\Big[(\vec d\!\cdot\!\vec k_\perp)\,A+(\vec d\!\cdot\!\vec u_\perp)\,B\Big],\nonumber\\
A&=&
-\frac{13\,t}{30\,\omega}
+\frac{\vec k_\perp^{\,2}}{6\,\omega\,\hat q_0}
-\frac{\vec k_\perp\!\cdot\!\vec u_\perp}{\hat q_0\,t}
+\frac{12}{5}\,\frac{\omega\,\vec u_\perp^{\,2}}{\hat q_0\,t^{2}} ,\nonumber\\
B&=&
-1+\frac{\vec k_\perp^{\,2}}{2\,\hat q_0\,t}
-\frac{12}{5}\,\frac{\omega\,(\vec k_\perp\!\cdot\!\vec u_\perp)}{\hat q_0\,t^{2}}
+6\,\frac{\omega^2\,\vec u_\perp^{\,2}}{\hat q_0\,t^{3}}, \nonumber  \\
\vec u_\perp &\equiv & \vec r_\perp-\frac{t}{2\omega}\,\vec k_\perp ;\nonumber
\end{eqnarray}

\begin{eqnarray}
    \delta f_v&=&f_0(\vec k_\perp,\vec r_\perp,t)\Bigl[(\vec v_\perp\cdot\vec k_\perp) C +(\vec v_\perp\cdot\vec r_\perp)D\Bigr],\nonumber\\
C&=& -\frac{1}{10\omega}
-\frac{8}{5}\frac{\vec k_\perp^{\,2}}{\hat q_0\,t\,\omega}
+\frac{22}{5}\frac{\vec k_\perp\cdot\vec r_\perp}{\hat q_0\,t^{2}}
-\frac{12}{5}\frac{\omega\,\vec r_\perp^{\,2}}{\hat q_0\,t^{3}}, \nonumber\\
D&=& \frac{6}{5t}
-\frac{14}{5}\frac{\vec k_\perp^{\,2}}{\hat q_0\,t^{2}}
+\frac{36}{5}\frac{\omega\,\vec k_\perp\cdot\vec r_\perp}{\hat q_0\,t^{3}}
-\frac{36}{5}\frac{\omega^{2}\vec r_\perp^{\,2}}{\hat q_0\,t^{4}} . \nonumber
\end{eqnarray}

Integrating the above expressions over the transverse area $\int d^2r_\perp$, one obtains the corrections to the transverse momentum distributions in Eqs~(\ref{eq:dfd}) and (\ref{eq:dfv}).  

The transverse momentum shift densities from these corrections are

\begin{eqnarray}
\delta \vec K_d &\equiv& \frac{1}{\hat q_0 t}
\int\frac{d^2k}{(2\pi)^2}\,\vec k_\perp\,k^2_\perp\, \delta f_d(\vec k_\perp,\vec r_\perp,t) \nonumber\\
&=& \frac{\omega r_0^2}{3t^2}f_0(\vec r_\perp,t)
\Bigg[
-\vec d\,t\Big(\frac{11}{320}+\frac{9r_\perp^2}{64r_0^2}+\frac{81\,r_\perp^4}{640r_0^4}\Big) \nonumber \\
& &+\frac{\vec r_\perp(t\,\vec d\!\cdot\!\vec r_\perp)}{r_0^2}\Big(\frac{9}{10}+\frac{27r_\perp^2}{80r_0^2}
+\frac{243r_\perp^4}{64r_0^4}\Big)
\Bigg], \nonumber
\end{eqnarray}

\begin{eqnarray}
\delta \vec K_v &\equiv& \frac{1}{\hat q_0 t}
\int\frac{d^2k_\perp}{(2\pi)^2}\,\vec k_\perp\,k^2_\perp\, \delta f_v(\vec k_\perp,\vec r_\perp,t) \nonumber\\
&=&  \frac{\omega r_0^2}{3t^2}f_0(\vec r_\perp,t)\left[-\vec v_\perp\left(\frac{39}{160}+\frac{45}{64}\frac{r_\perp^2}{r_0^2}-\frac{81}{160} \frac{r_\perp^4}{r_0^4}\right) \right. \nonumber\\
&& -\left. \frac{\vec r\,(\vec v_\perp\!\cdot\!\vec r_\perp)}{r_0^2}\left(\frac{891}{160}+\frac{1701}{160}\frac{r_\perp^2}{r_0^2}+\frac{729}{40}\frac{r_\perp^4}{r_0^4}\right)\right], \nonumber
\end{eqnarray}
where
$$
r_0^2=\frac{\hat q_0 t^3}{\omega^2},\,\, f_0(\vec r_\perp,t)=\frac{3}{\pi r_0^2} \exp\left(-\frac{3r_\perp^2}{r_0^2}\right).
$$

\begin{figure}
\centerline{\includegraphics[width=8.0cm]{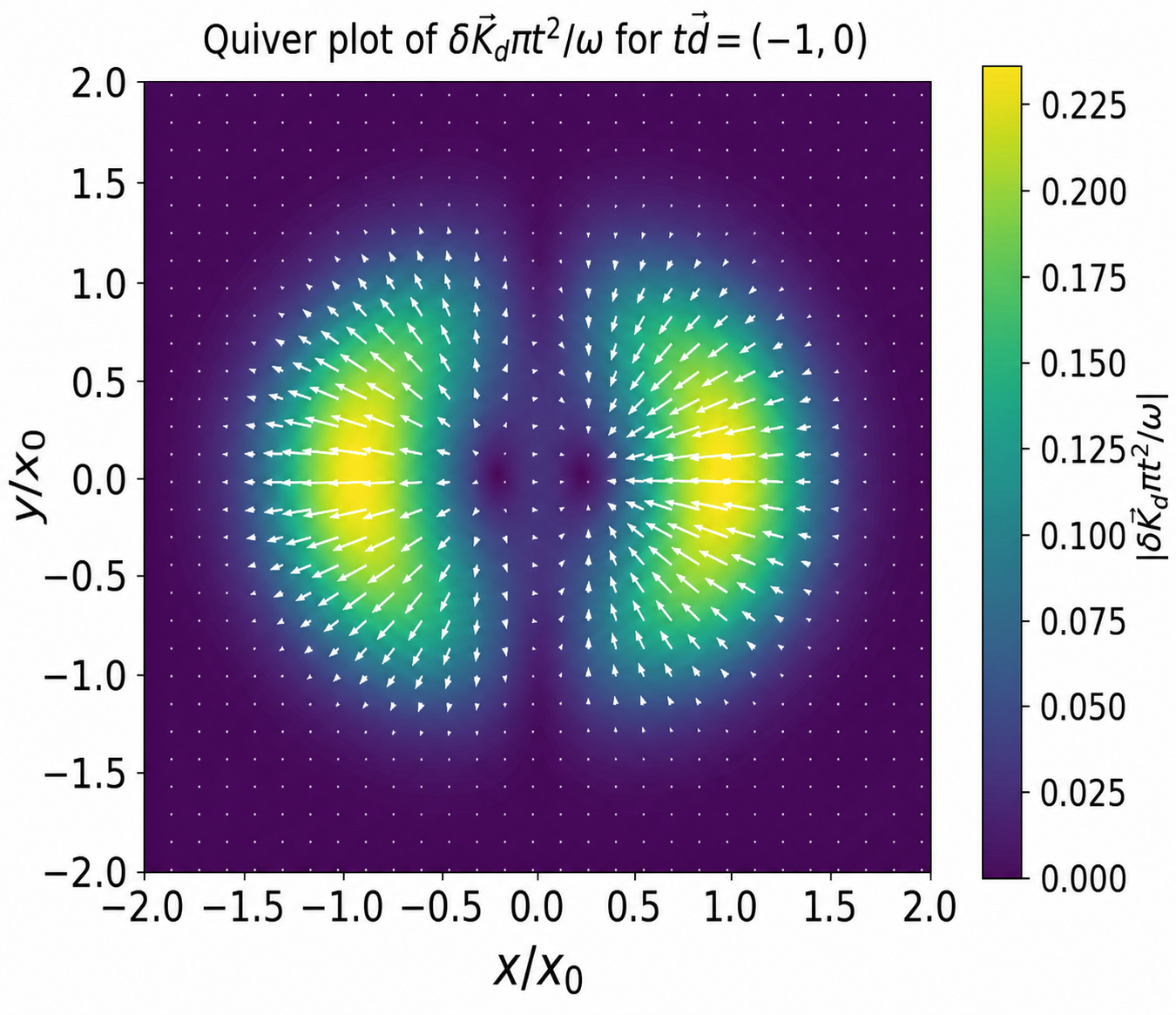}}
 \caption{ Quiver plot of $\delta \vec K_d$ (color scale for the amplitude and arrows for the direction) for $\vec d t=(-1,0)$.} 
 \label{eq:kd}
\end{figure}

\begin{figure}
\centerline{\includegraphics[width=8.0cm]{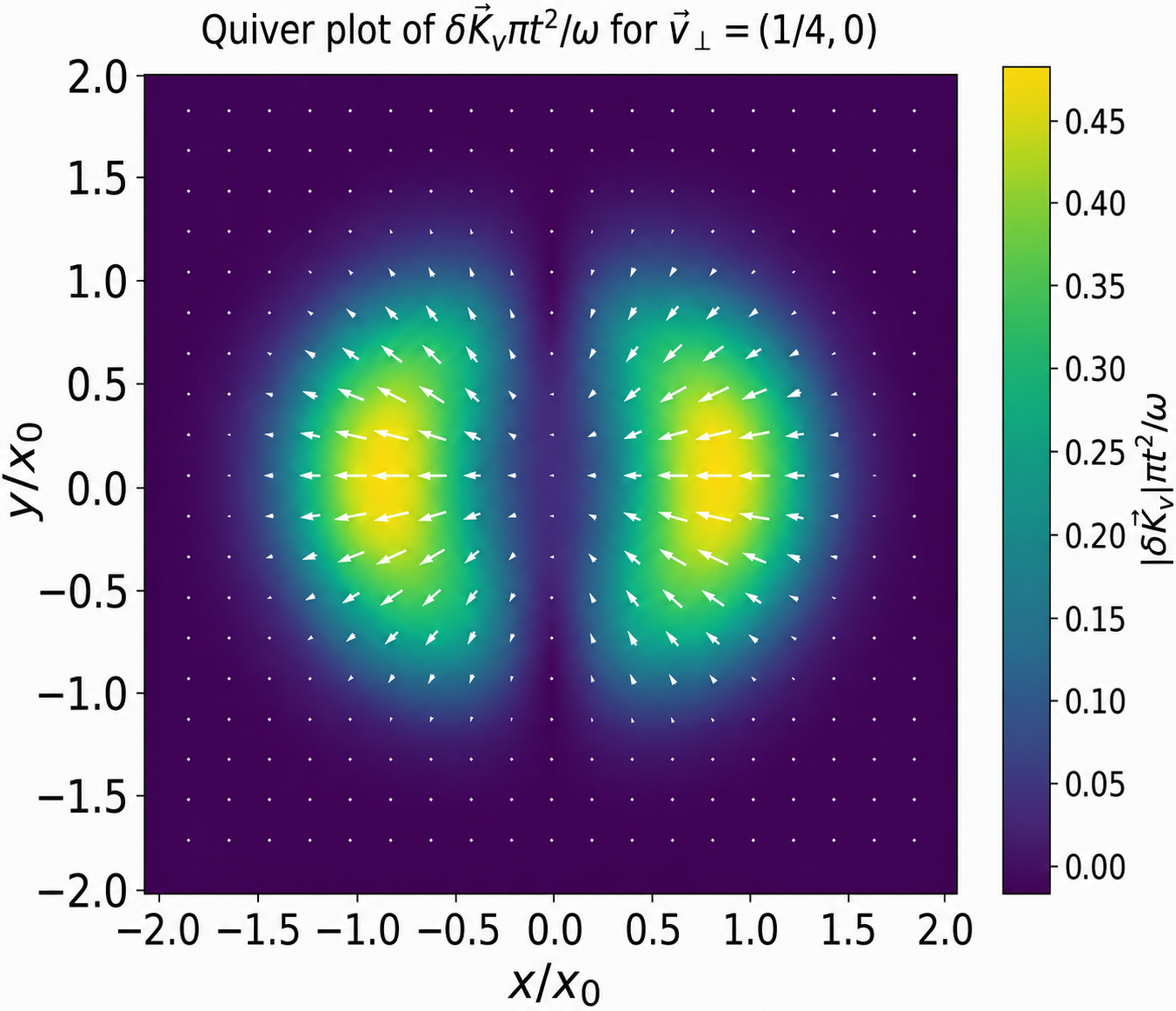}}
 \caption{ Quiver plot of $\delta \vec K_v$ (color scale for the amplitude and arrows for the direction) for $\vec v_\perp=(1/4,0)$.} 
 \label{eq:kv}
\end{figure}

Figs.~\ref{eq:kd} and \ref{eq:kv} show the quiver plots of the transverse-momentum-shift distributions $\delta \vec K_d$ and $\delta \vec K_v$ where the color scale indicates the magnitude and the arrows indicate the direction.

Integrating over the transverse area of the above, one obtains the total transverse-momentum shifts in Eq.~(\ref{eq:shift}).
\\

\begin{figure}
\centerline{\includegraphics[width=8.0cm]{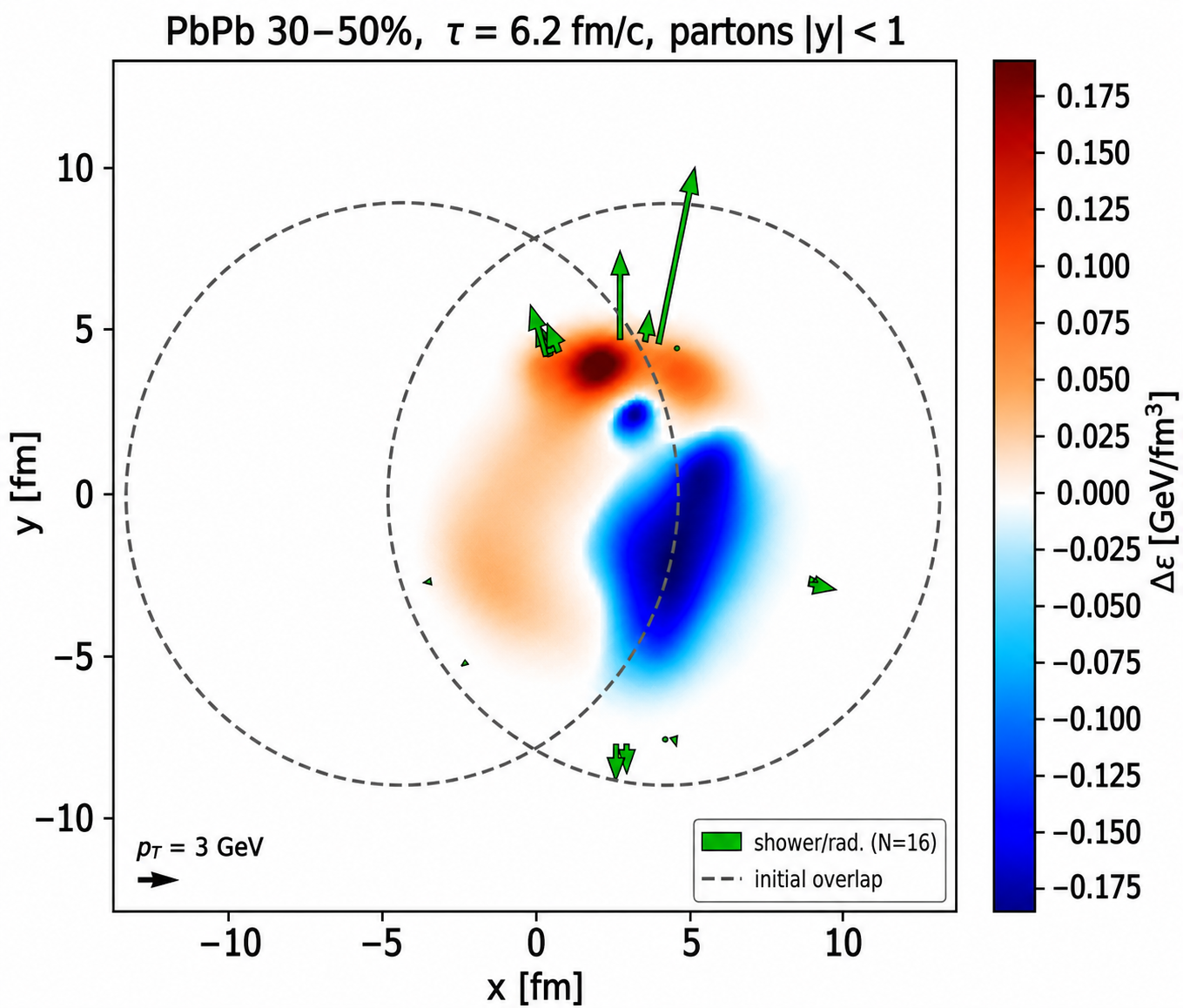}}
 \caption{A snapshot of jet-induced
energy density distribution in the transverse plane of a semi-central 30-50\% Pb+Pb collision with a $\gamma$-jet at the spatial rapidity $|\eta_s|<1.0$
and proper time $\tau$= 6.2 fm/c from CoLBT-hydro simulations. Lines with arrows represent
the transverse momenta of partons and dashed circles represent the two colliding nuclei.}
\label{fig:density}
\end{figure}

\noindent {\bf 2: Energy density distribution of jet-induced medium response:}
Shown in Fig.~\ref{fig:density} is a snapshot of the energy density distribution (color scale) of jet-induced medium response by a $\gamma$-jet and hard partons (lines with arrows)  at time $\tau=6.2$ fm after the initial jet production from CoLBT-hydro simulations. One can see that the medium response moves towards the dense region of the overlap while the diffusion wake moves away. One can also see a second medium response generated by another hard parton at a later stage.

\begin{figure}
    \centering
    \includegraphics[width=0.9\linewidth]{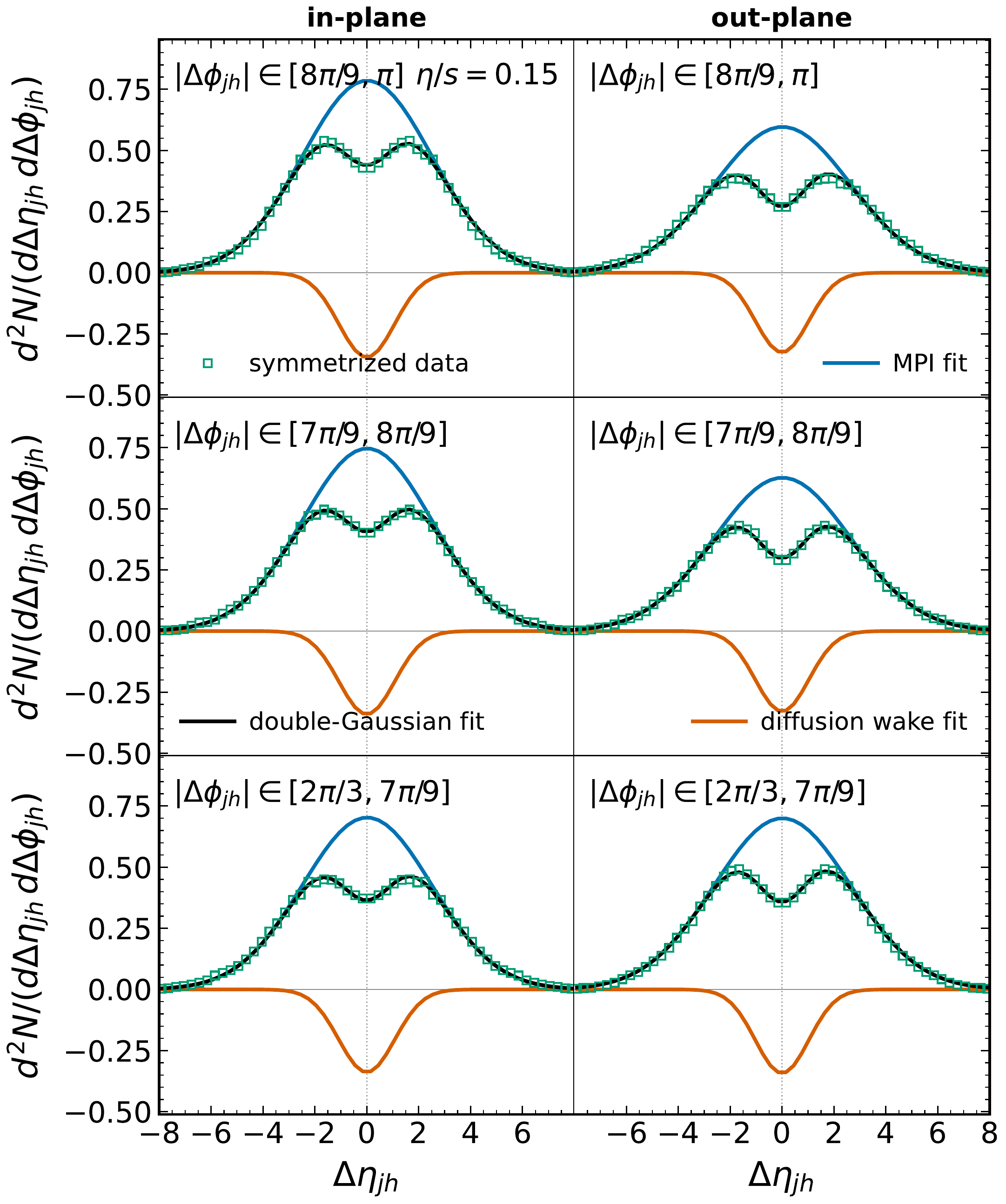}
    \caption{Double-Gaussian fits to the jet-hadron correlations (black lines) in rapidity 
from CoLBT-hydro simulations (open squares) of in-plane (left) and out-plane (right) $\gamma$-jets in 30-50\% Pb+Pb collisions at $\sqrt{s}=5.02$ TeV for three different bins of the azimuthal angle $\Delta\phi_{\rm jh}$ with MPI-ISR (blue lines) and diffusion wake (red lines) contributions.}
    \label{fig:doublegaussian}
\end{figure}

\begin{figure}[H]
    \centering
    \includegraphics[width=0.99\linewidth]{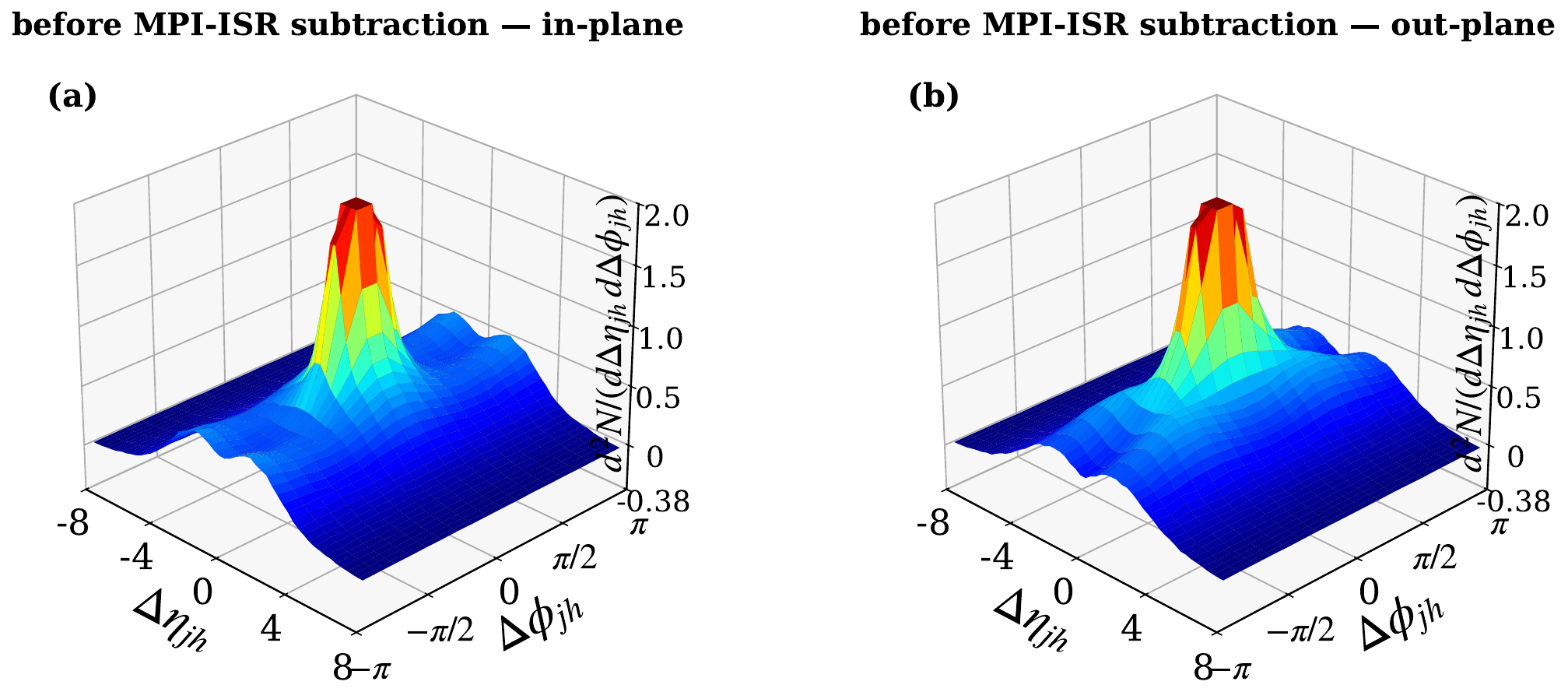}
    \includegraphics[width=0.99\linewidth]{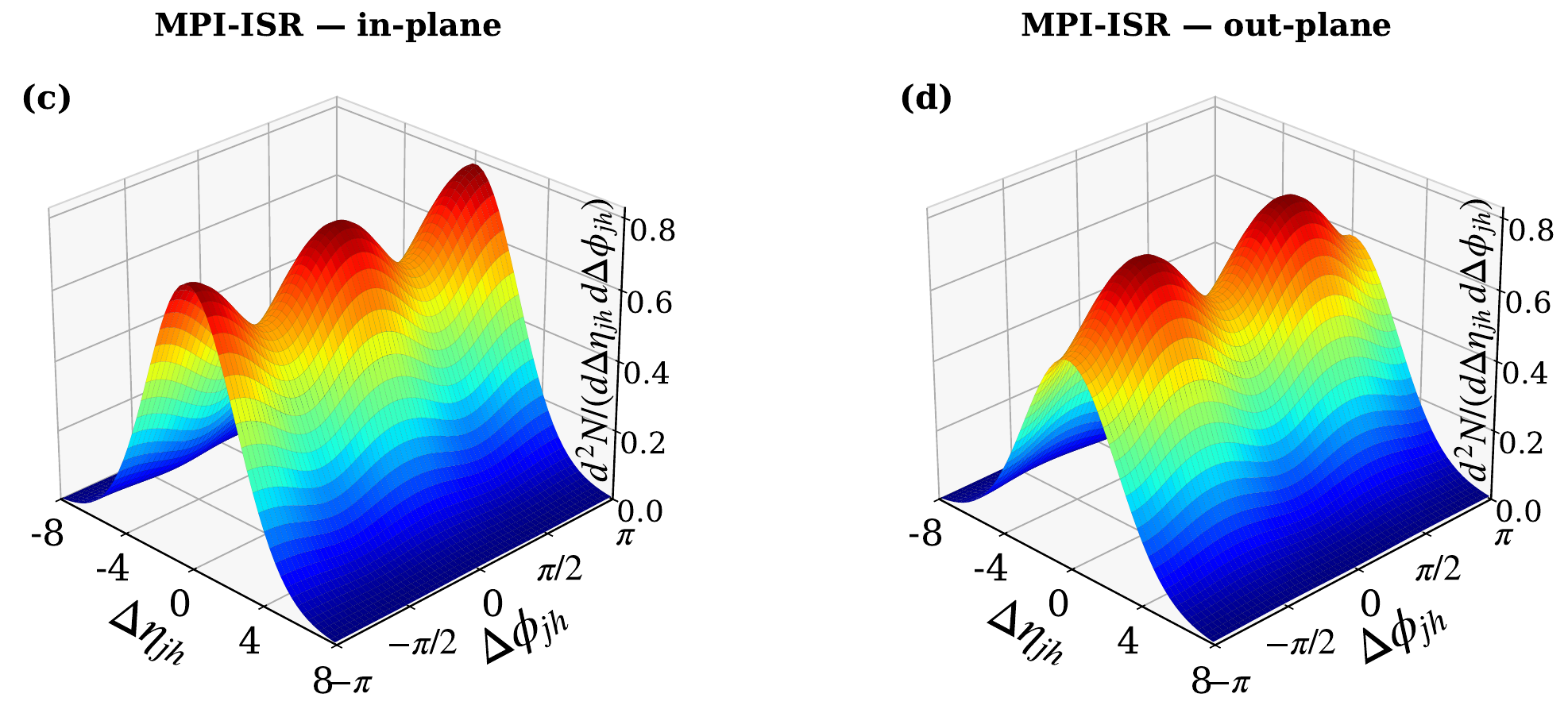}
    \includegraphics[width=0.99\linewidth]{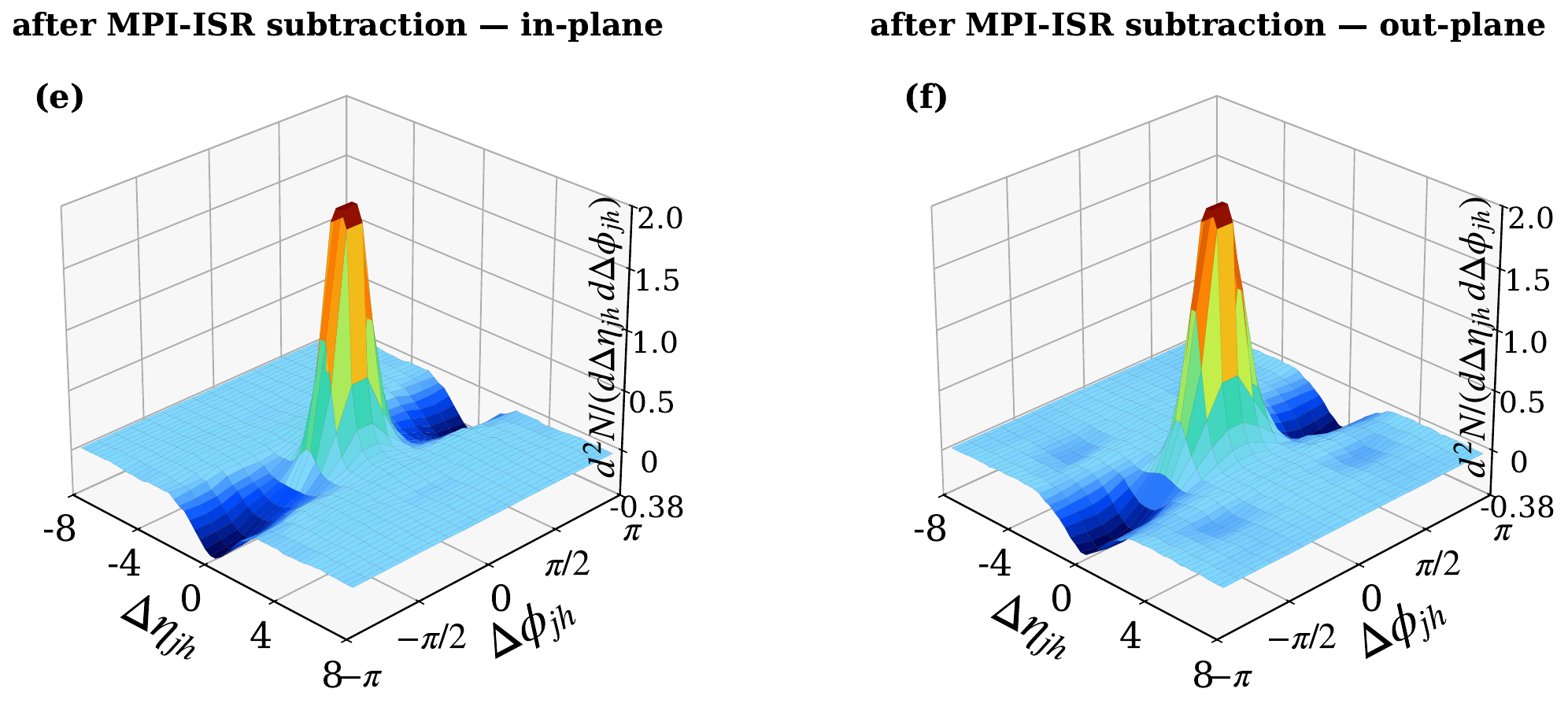}
    \includegraphics[width=0.99\linewidth]{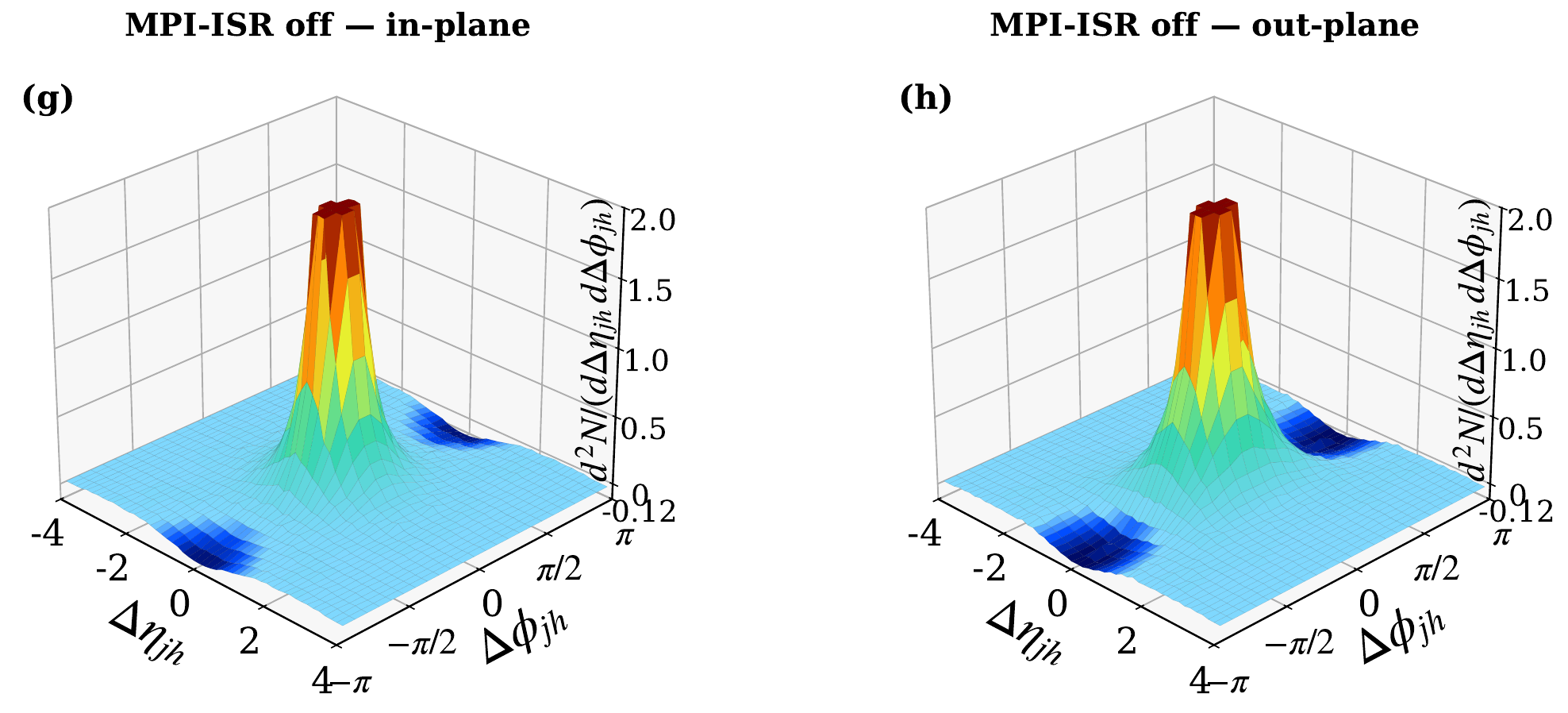}
    \caption{2D jet-hadron correlations (a and b) in (left) in-plane and (right) out-plane $\gamma$-jets events of 30-50\% collisions at $\sqrt{s}=5.02$ TeV, the MPI-ISR contributions from double-Gaussian fits (c and d), the jet-hadron correlations after the subtraction of MPI-ISR contributions (e and f), and the jet-hadron correlation from CoLBT-hydro simulations with MPI-ISR turned off (g and h).}
    \label{fig:2Dplots}
\end{figure}

\noindent {\bf 3: Double-Gaussian fit of the jet-hadron correlation in rapidity:}
Using double-Gaussian fits of the jet-hadron correlation in rapidity, we can separate the MPI-ISR contribution and diffusion wake at three different bins of the azimuthal angle $\Delta\phi_{\rm jh}$ in the $\gamma$ direction as shown in Fig.~\ref{fig:doublegaussian}. We also show  in Fig.~\ref{fig:2Dplots} the 2D jet-hadron correlations (a,b), the MPI-ISR contributions (c,d), the jet-hadron correlations after the subtraction of MPI-ISR contributions (e,f), the jet-hadron correlation from CoLBT-hydro simulation with MPI-ISR turned off (g,h) for in-plane (left) and out-plane (right) $\gamma$-jets.

\end{document}